# HEAD-TAIL MODES FOR STRONG SPACE CHARGE

A. Burov

FNAL*, Batavia, IL 60510, U.S.A.

*Abstract*
The head-tail modes are described for the space charge tune shift significantly exceeding the synchrotron tune. A general equation for the modes is derived. The spatial shapes of the modes, their frequencies, and coherent growth rates are explored. The Landau damping rates are also found. The suppression of the transverse mode coupling instability by the space charge is explained.

## 1. INTRODUCTION

The head-tail instability of bunched beams was observed and theoretically described many years ago [1-3]. Since then, this explanation has been accepted and included in textbooks [4,5], but still there is an important gap in the theory of head-tail interaction. This relates to the influence of space charge on the coherent modes: their shapes, growth rates and Landau damping. A significant theoretical paper on that issue was published by M. Blaskiewicz ten years ago [6]. In particular, a compact analytical description of the coherent modes was found there for a square well model, air-bag distribution and a short-range wake function, without any assumption for the relative values of the space charge tune shift, the synchrotron tune and the coherent tune shift. Later, the air-bag limitation was removed for zero-wake case in the square well [7]. Ref. [6,7] shed the first light on the problem, reminding us that because of the very specific restrictions of that model, of all the questions that were yet to be answered. Here, an attempt to provide answers to those questions is presented.

Compared to the work reported in Ref. [6,7], this attempt is both broader and narrower. It is broader since there are no assumptions about the shape of the potential well, the bunch distribution function, and the wake function. The solution for a parabolic potential well and 3D Gaussian bunch is given in detail, but the method is universal. Since this paper deals with arbitrary distribution functions, the Landau damping is not generally zero; it is calculated in this paper.

From another aspect, though, my approach is narrower than that of Ref. [6], since a certain condition between the important parameters is assumed below. Namely, it is assumed that the space charge tune shift in the bunch 3D center $Q_{max}$ is large compared to both to the synchrotron tune $Q_s$ and the wake-driven coherent tune shift $Q_w$: $Q_{max} >> Q_s, Q_w$.

The structure of this paper is as follows. In the next section, a single-particle equation of motion is written in the rigid-beam approximation and the validity of this approximation is discussed. Then the coherent modes are found from this equation for the square potential well with an arbitrary beam distribution function. The vanishing Landau damping for the square well is pointed out.

After that, the problem for the coherent modes in the presence of strong space charge is reduced to a second-order ordinary differential equation with zero boundary conditions for the eigenfunction derivative; this is done for an arbitrary beam distribution function and potential well. The eigen-modes and eigen-frequencies are found for the no-wake case, when only the space charge and the synchrotron motion are taken into account. At this step, the mode shapes $\bar{y}_k(\tau)$ and frequencies $\nu_k$ are found and the wake-driven coherent growth rates are calculated as perturbations. When the mode structure is described both in general and in detail for the Gaussian bunch, the Landau damping rates $\Lambda_k$ are found in the next two sections: first, without and second, with transverse nonlinearity.

In the last section of this paper, the limit of weak head-tail is removed, and the transverse mode coupling instability (TMCI) at strong space charge is discussed. It is shown that in this case, the TMCI threshold typically exceeds its naïve estimate based on the mode separation by a factor of 10-100.

## 2. RIGID BEAM EQUATIONS

In order to keep the same notation as in Ref. [6], let $\theta$ be the time in radians, $\tau$ - a distance along the bunch in radians as well, $X_i(\theta)$ - a betatron offset of *i*-th particle, and $\bar{X}(\theta, \tau)$ - an offset of the beam center at the given time $\theta$ and position $\tau$. Since all the tune shifts are small compared with the bare betatron tune $Q_b$, the latter can be excluded from the considerations by using slow variables $x_i(\theta)$:

$$X_i(\theta) = \exp(-iQ_b\theta)x_i(\theta).$$

After that, a single-particle equation of motion can be written as

$$\dot{x}_i(\theta) = iQ(\tau_i(\theta))[x_i(\theta) - \bar{x}(\theta, \tau_i(\theta))] - i\varsigma\nu_i(\theta)x_i(\theta) - i\kappa\hat{\mathbf{W}}\bar{x}. \quad (1)$$

---

Here a dot over a variable $x_i$ stands for the time derivative, $\dot{x}_i = dx_i/d\theta$; the effective chromaticity $\zeta = -\xi/\eta$ with $\xi = dQ_b/d(\Delta p/p)$ as the conventional chromaticity and $\eta = \gamma_t^{-2} - \gamma^{-2}$ as the slippage factor; $Q(\tau)$ is the space charge tune shift as a function of the position inside the bunch; $v_i(\theta) = \dot{\tau}_i(\theta)$ is the velocity within the bunch, and $\kappa\hat{\mathbf{W}}\bar{x}$ is the wake force expressed in terms of the wake linear operator $\hat{\mathbf{W}}$ to be specified below.

Note that Eq. (1) already assumes a rigid-beam approximation. This approximation is based on the idea that the transverse coherent motion of the beam can be treated as displacements of beam longitudinal slices, so the force on a given particle is just proportional to its offset from the local beam centroid. For a coasting beam, the validity of the rigid-beam model is discussed in Ref. [8]. To be justified, the rigid-beam model requires a sufficient separation between the coherent frequency and the incoherent spectrum: the separation has to be significantly larger than the width of the bare incoherent spectrum. As a result almost all the particles respond almost identically to the collective field.

The chromaticity term can be excluded from Eq. (1) with a substitution $x_i(\theta) = y_i(\theta)\exp(-i\zeta\tau_i(\theta))$, leading to

$$\dot{y}_i(\theta) = iQ(\tau_i(\theta))[y_i(\theta) - \bar{y}(\theta,\tau_i(\theta))] - i\kappa\hat{\mathbf{W}}\bar{y} \quad (2)$$

with

$$\kappa = \frac{r_0 R}{4\pi\beta^2\gamma Q_b};$$

$$\hat{\mathbf{W}}\bar{y} = \int_\tau^\infty W(\tau-s)\exp(i\zeta(\tau-s))\rho(s)\bar{y}(s)ds. \quad (3)$$

Here $r_0$ is the classical radius of the beam particles; $R=C/(2\pi)$ is the average ring radius; $\beta$ and $\gamma$ are the relativistic factors, $\rho(s)$ is the bunch linear density normalized on the number of particles in the bunch, $\int\rho(s)ds = N_b$, and the wake-function $W(s)$ is defined according to Ref [4] (slightly different from the definition of Ref. [6]).

We begin from solving Eq. (2) for the no-wake case. Next the wake is taken into account as a perturbation of the space charge eigen-modes. These unperturbed eigen-modes are to be found from the no-wake reduction of Eq. (2):

$$\dot{y}_i(\theta) = iQ(\tau_i(\theta))[y_i(\theta) - \bar{y}(\theta,\tau_i(\theta))] \quad (4)$$

Solutions of this equation give the space charge eigen-modes: their spatial shapes and frequencies. The modes do not depend on the chromaticities, except for the common head-tail phase factor $\exp(-i\zeta\tau)$. The chromaticity enters into the problem through the wake term, Eq. (3), affecting the coherent growth rates. As it will be seen below, the chromaticity normally makes the coherent growth rates negative for the modes, which number $k$ is smaller than the head-tail phase, $k \lesssim \zeta\sigma$, with $\sigma$ as the rms bunch length.

## 3. SQUARE POTENTIAL WELL

Before going to a general case, it would be instructive to solve the easier problem of the square well potential. Similar to Ref. [7], this problem is solved for a general distribution over the synchrotron frequencies, and for no-wake case. When the wake field is small enough, $Q_w \ll \min(Q_s^2/Q_{\max}, Q_s)$, the solution is extended to the weak head-tail case. In this section only, the ratio between the space charge tune shift and the synchrotron tune can be arbitrary.

For no-wake case, a single-particle equation (4) has a constant coefficient $Q(\tau) = Q$, and can be easily solved:

$$y_i(\theta) = -iQ\int_{-\infty}^\theta \bar{y}(\theta',\tau_i(\theta'))\exp(iQ(\theta-\theta'))d\theta'. \quad (5)$$

To find the eigen-modes, the boundary conditions for the beam centroid have to be taken into account. Since every particle is reflected instantaneously from the walls of the potential well, its offset derivative cannot immediately change after the reflection. This, in turn, leads to a conclusion that a space derivative of the beam centroid is zero at the bunch boundaries $\tau = 0$ and $\tau = l$:

$$\left.\frac{\partial}{\partial\tau}\bar{y}(\theta,\tau)\right|_{\tau=0} = \left.\frac{\partial}{\partial\tau}\bar{y}(\theta,\tau)\right|_{\tau=l} = 0. \quad (6)$$

Thus, the centroid offset can be Fourier-expanded as

$$\bar{y}(\theta,\tau) = \exp(-iv_k\theta)\sum_{m=0}^\infty C_m\cos(\pi m\tau/l), \quad (7)$$

with $C_m$ as yet unknown coefficients, and $v_k$ as the eigen-frequencies to be found. For the right-hand side of single-particle Eq. (5), it gives

$$\bar{y}(\theta',\tau_i(\theta')) = \exp(-iv_k\theta')\sum_{m=0}^\infty C_m\cos\left(\pi m\frac{\tau_i(\theta) - v_i(\theta-\theta')}{l}\right). \quad (8)$$

Substituting Eq. (8) into Eq. (5) and after taking the integral, we obtain the result

$$y_i(\theta) = \exp(-iv_k\theta)\sum_{m=0}^\infty C_m\cos\left(\pi m\frac{\tau_i(\theta)}{l}\right)\frac{Q(Q+v_k)}{(Q+v_k)^2 - m^2 Q_{si}^2}, \quad (9)$$

with $Q_{si} = \pi v_i/l$ as the synchrotron frequency of the *i*-th particle. Averaging Eq. (9) over all the particles at the given location yields the shape of the eigen-modes as

$$\bar{y}_k(\theta,\tau) = \exp(-iv_k\theta)\bar{y}_k(\tau); \quad \bar{y}_k(\tau) = \sqrt{2/l}\cos(\pi k\tau/l). \quad (10)$$

These eigen-modes constitute a full orthonormal basis:

$$\int_0^l \bar{y}_k(\tau)\bar{y}_m(\tau)d\tau = \delta_{km}. \quad (11)$$

The coherent shifts $v_k$ have to be found from the following dispersion equation:

$$1 = \int\frac{Q(Q+v_k)f(Q_s)dQ_s}{(Q+v_k+i0)^2 - k^2 Q_s^2}; \quad k = 0,1,2... \quad (12)$$

Here, the Landau rule $v_k \to v_k + i0$ was taken into account, and the distribution function over the synchrotron frequencies is assumed to be normalized:

$$\int f(Q_s)dQ_s = 1.$$

The dispersion equation (12) for the square well model and arbitrary longitudinal distribution was obtained by Blaskiewicz [7]. The simplest case for the dispersion equation (12) is the air-bag distribution, fully considered in Ref. [6]. Taking $f(Q_s) = \delta(Q_s - \overline{Q}_s)$, the result of Ref. [6] is reproduced:

$$v_k = -\frac{Q}{2} \pm \sqrt{\frac{Q^2}{4} + k^2 \overline{Q}_s^2} \quad (13)$$

Note that the positive and negative frequencies are significantly different. If the synchrotron tune is so small that $k\overline{Q}_s \ll Q/2$, the negative eigen-frequencies almost coincide with the singe-particle tune shift, while the positive ones are much smaller than that. The negative solution $v_k \approx -Q$ is almost equal to the single-particle tunes, so this mode is easily Landau-damped, and thus is excluded from the further analysis. The other solution yields in this case

$$v_k = k^2 \overline{Q}_s^2 / Q \quad . \quad (14)$$

Note that instead of zero space charge spectrum $Q_k = k\overline{Q}_s$, the mode frequency here is quadratic with its mode number.

When there is some synchrotron frequency spread, it is possible for some particles to be resonant with the mode, providing the Landau damping. For them, the denominator in the dispersion equation goes to zero, so their synchrotron tunes are

$$kQ_s = \pm(Q + v_k) \approx \pm Q \quad . \quad (15)$$

For typical practical cases, where the synchrotron tune is order(s) of magnitude smaller, than the maximal space charge tune shift, this condition selects very distant tails, so we conclude that as a practical matter there is no Landau damping here.

We have not, up to this point, taken into account the nonlinear betatron tune shift, This tune shift $\delta Q(J_1, J_2)$, being a function of the two transverse actions $J_1, J_2$, modifies the dispersion relation similarly to the coasting beam case, resulting in

$$1 = -\int \frac{\partial f}{\partial J_1} \frac{J_1 Q(Q + v_k) dQ_s dJ_1 dJ_2}{(Q - \delta Q(J_1, J_2) + v_k + i0)^2 - (kQ_s)^2}, \quad (16)$$

with the normalized distribution function $\int f(Q_s, J_1, J_2) dQ_s dJ_1 dJ_2 = 1$ assuming that we are studying oscillations along the 1$^{st}$ degree of freedom. The dispersion relation (16) is valid for any dependence of the space charge tune shift on the transverse actions $Q \rightarrow Q(J_1, J_2)$. When the space charge tune shift is much larger than the synchrotron tune and the nonlinear tune shift, the solution of the dispersion equation (16) follows:

$$v_k = k^2 \frac{\int \frac{fQ_s^2 dQ_s dJ_1 dJ_2}{Q^2(J_1, J_2)}}{\int \frac{fdQ_s dJ_1 dJ_2}{Q(J_1, J_2)}} + 2 \frac{\int \frac{f\delta Q(J_1, J_2) dQ_s dJ_1 dJ_2}{Q(J_1, J_2)}}{\int \frac{fdQ_s dJ_1 dJ_2}{Q(J_1, J_2)}}$$

Since this solution does not assume any special relation between the space charge tune shift and the synchrotron tune, it is worthwhile to look at the case of relatively low space charge, $Q_s \gg Q/2$. As it is clear from Eq. (13), the collective modes are separated from the incoherent spectrum by one-half of the space charge tune shift. Without transverse nonlinearity, the Landau damping is provided by the synchrotron tune spread. The integrand of the dispersion equation (12) has its poles at $kQ_s = Q + v_k$. With the eigenvalue (13), it yields that the Landau damping is provided by particles whose synchrotron tune $Q_s$ deviates from the average synchrotron tune $\overline{Q}_s$ by $Q_s - \overline{Q}_s = Q/(2k)$. For $k = 0$, there is no Landau damping without transverse nonlinearity, since $v_0 = 0$ is an exact solution of the dispersion equation (12), similar to the coasting beam case. Note that for this low space charge case, the rigid-beam approximation is valid for both signs in Eq. (13), as soon as all the resonant particles are located only in tails of the distribution.

It is worth to note that the square well model greatly underestimates the Landau damping. The reason is that in case of the square well, there is no spatial variation of the space charge tune shift. However, for realistic buckets and bunch shapes, the space charge tune shift smoothly drops to zero at the longitudinal tails, making possible the wave-to-particle energy transfer there. This appears to be a leading mechanism of Landau damping for bunched beams, missing in the square well case, and being considered in detail in two sections below.

To finish our analysis of the square well model, there is one more thing to do.. After the set of the eigen-modes is found for zero-wake case, the wake can be taken into account by means of perturbation theory, assuming it is small enough. This step is relatively simple. Indeed, the wake term in Eq. (2) causes its own tune shift $Q_w$, leading to additional factor $\exp(-iQ_w \theta)$ for the single-particle offset on the left-hand side of Eq. (2). This immediately turns this equation into $Q_w y_i = \hat{\mathbf{W}} \bar{y}$. After averaging, this gives the wake tune shift as a diagonal element of the wake operator:

$$Q_w = (\bar{y}, \hat{\mathbf{W}}\bar{y}) \equiv \kappa \int\limits_{-\infty}^{\infty}\int\limits_{\tau}^{\infty} W(\tau - s) \exp(i\zeta(\tau - s)) \rho(s) \bar{y}_k(s) \bar{y}_k(\tau) ds d\tau. \quad (17)$$

Here, the normalization (11) of the orthogonal modes $\bar{y}_k(\tau)$ (Eq. 10) was taken into account.

If the vacuum chamber is not round, the detuning, or quadrupole wake $D(\tau)$ modifies the coherent tune shifts [9]. Assuming a force from a leading particle (subscript 1) on a trailing particle (subscript 2) as $\propto W(\tau)x_1 + D(\tau)x_2$, it yields for the detuning coherent shift

$$Q_d = \kappa \int\limits_{-\infty}^{\infty}\int\limits_{\tau}^{\infty} D(\tau - s) \rho(s) \bar{y}_k^2(\tau) ds d\tau. \quad (18)$$

Instead of the wake tune shift (17), the detuning one in (18) is purely real; it does not affect the beam stability. This conclusion though is limited by the weak head-tail approximation, where the wake is so small that it can be taken as perturbation, leading to (17, 18). It was shown in Ref. [9], that it is not valid for the transverse mode coupling instability (TMCI), where the detuning wake normally increases the intensity threshold.

Note that the derivation of Eqs. (17, 18) does not use any specific features of the square well model; thus, these results are valid for any potential well and bunch profile, when a corresponding orthonormal basis of the eigen-modes is used.

Growth rates as functions of the head-tail phase $\zeta l$ are presented in Fig. 1 for the square well model with a constant wake function, $W(\tau) = W_0 = $const. The rates are given in units of $\kappa N_b W_0$ with $N_b = \rho l$ as a number of particles in the bunch.

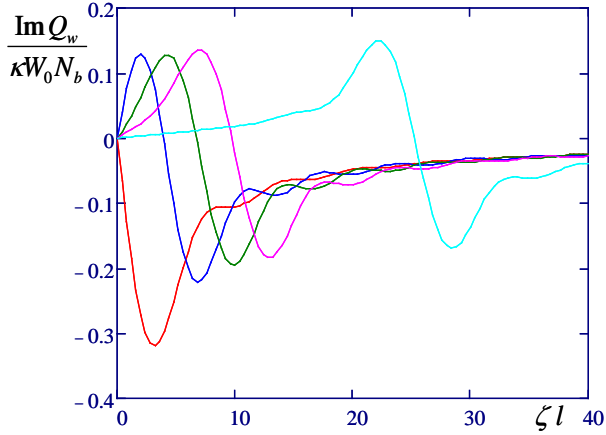

Fig. 1: Weak head-tail growth rates for the square potential well of the lowest mode (mode 0, red), mode 1 (blue), 2 (green), 3 (magenta) and 8 (cyan) as functions of the head-tail phase $\zeta l$ for a constant wake function. The rates are in units of $\kappa N_b W_0$.

## 4. GENERAL SPACE CHARGE MODES

In this section, an ordinary differential equation for the eigen-modes is derived for a general potential well and 3D bunch distribution function, assuming strong space charge,

$$Q \gg 2k\overline{Q}_s. \quad (19)$$

The fact that the bunch modes are described by single-argument functions, dependent on the position along the bunch only, is due to a strong coherence, introduced by the space charge. Indeed, the classical no-space-charge head-tail modes are generally described by their dependence both on the synchrotron phase and the synchrotron action, so the synchrobetatron modes are characterized by what are called azimuthal and radial numbers. All the radial modes are degenerate, having the same coherent tune, determined by the azimuthal number. With the strong space charge, all the individual degrees of freedom are detuned from the coherent motion by approximately the same number, namely, the local space charge tune shift. As a result, locally all the particles are moving almost identically; their position in the synchrotron phase space does not play a role, as soon as they are at the given longitudinal position. That is why the space charge modes are described by single-argument functions dependent on the position along the bunch only.

The single-particle equation (4) can be solved in general:

$$y_i(\theta) = -i \int_{-\infty}^{\theta} Q(\tau_i(\theta')) \overline{y}(\theta', \tau_i(\theta')) \exp(i\Psi(\theta) - i\Psi(\theta')) d\theta';$$

$$\Psi(\theta) \equiv \int_0^\theta Q(\tau_i(\theta')) d\theta' \quad. \quad (20)$$

Since $Q(\tau) > 0$, $\Psi(\theta)$ is monotonic and so integration over $\theta$ in Eq. (20) can be replaced by integration over $\Psi$:

$$y_i(\Psi) = -i \int_{-\infty}^{\Psi} \overline{y}(\Psi') \exp(i\Psi - i\Psi') d\Psi'. \quad (21)$$

Note that due to (19), the phase $\Psi$ runs fast compared with relatively slow dependence $\overline{y}(\Psi)$, so the later can be expanded in a Taylor series-,

$$\overline{y}(\Psi') \approx \overline{y}(\Psi) - (\Psi - \Psi') \frac{d\overline{y}}{d\Psi} + \frac{(\Psi - \Psi')^2}{2} \frac{d^2\overline{y}}{d\Psi^2} \quad.$$

After that the integral is easily evaluated:

$$y_i(\Psi) = \overline{y}(\Psi) - i\frac{d\overline{y}}{d\Psi} - \frac{d^2\overline{y}}{d\Psi^2} \quad. \quad (22)$$

To come back to original variables, one can use that

$$\frac{d}{d\Psi} = \frac{\mathrm{v}}{Q(\tau)} \frac{\partial}{\partial \tau} + \frac{1}{Q(\tau)} \frac{\partial}{\partial \theta} = \frac{1}{Q(\tau)} \left( \mathrm{v} \frac{\partial}{\partial \tau} - i\nu \right). \quad (23)$$

Applied to Eq. (22), this gives

$$y_i = \overline{y} - \frac{i}{Q(\tau)} \left( \mathrm{v} \frac{\partial}{\partial \tau} - i\nu \right) \overline{y} - \left[ \frac{1}{Q(\tau)} \left( \mathrm{v} \frac{\partial}{\partial \tau} - i\nu \right) \right]^2 \overline{y}. \quad (24)$$

At this point, we can average over velocities v at the given position $\tau$. Doing this, the eigenvalue $\nu_k$ can be neglected in the second-order term of Eq. (24), supposing $|\nu| \ll |\mathrm{v}\partial/\partial \tau|$, as it is true for the square well bucket and is confirmed below for general case. After that, the equation for eigen-modes follows as a second-order ordinary self-adjoint differential equation:

$$\nu \overline{y} + u(\tau) \frac{d}{d\tau} \left( \frac{u(\tau)}{Q(\tau)} \frac{d\overline{y}}{d\tau} \right) = 0;$$

$$u^2(\tau) \equiv \frac{\int_{-\infty}^{\infty} \mathrm{v}^2 f(\mathrm{v}, \tau) d\mathrm{v}}{\int_{-\infty}^{\infty} f(\mathrm{v}, \tau) d\mathrm{v}}, \quad (25)$$

where $f(\mathrm{v},\tau)$ is a normalized steady-state longitudinal distribution function, $f(\mathrm{v},\tau) = f(H(\mathrm{v},\tau))$, with $H(\mathrm{v},\tau)$ as the longitudinal Hamiltonian.

It has to be noted, that the derivation of Eq. (25) from Eq. (24) implicitly assumed that the space charge tune shift depends only on the longitudinal position, and does not depend on the individual transverse amplitude. It is possible, however, to remove this limitation, and to see that Eq. (25) is actually valid for any transverse distribution, after certain redefinition of the space charge tune shift $Q(\tau)$. Indeed, the single-particle Eq. (24) does not make any assumption about the individual space charge tune shift dependence $Q(\tau)$, which can be considered as dependent on the transverse actions $J_{1i}, J_{2i}$ as well: $Q(\tau) \to Q_i(\tau) = Q(J_{1i}, J_{2i}, \tau)$. The averaging of Eq. (24) just has to take into account this dependence of the space charge tune shift on the transverse actions. As an example, for a Gaussian round beam, i.e. a beam with identical emittances and beta-functions, the transverse dependence of the space charge tune shift can be calculated as [5,8]:

$$Q(J_1, J_2, \tau) = Q_{\max}(\tau) \int_0^1 \frac{\left[ I_0\left(\frac{J_1 z}{2}\right) - I_1\left(\frac{J_1 z}{2}\right) \right] I_0\left(\frac{J_2 z}{2}\right)}{\exp(z(J_1 + J_2)/2)} dz \equiv \quad (26)$$
$$\equiv Q_{\max}(\tau) g(J_1, J_2)$$

Here $J_1, J_2$ are two dimensionless transverse actions, conventionally related to the offsets as $x = \sqrt{2 J_1 \varepsilon_1 \beta_1} \cos(\psi)$ with $\varepsilon_1$ and $\beta_1$ as the rms emittance and beta-function, so that the transverse distribution function is

$$f_\perp(J_1, J_2) = \exp(-J_1 - J_2). \quad (26a)$$

The transverse averaging of Eq. (24) requires calculation of two transverse moments $q_{-1}, q_{-2}$ of the tune shift $Q(J_1, J_2, \tau)$ generally defined by:

$$\left\langle Q_i^p(\tau) \right\rangle_\perp = \int_0^\infty \int_0^\infty dJ_1 dJ_2 f_\perp(J_1, J_2) Q^p(J_1, J_2, \tau) \equiv q_p^p Q_{\max}^p(\tau).$$
$$q_p = \left[ \int_0^\infty \int_0^\infty dJ_1 dJ_2 f_\perp(J_1, J_2) g^p(J_1, J_2) \right]^{1/p} \quad (27)$$

After that, Eq. (25) follows for any transverse distribution with a substitution

$$Q(\tau) \to Q_{\text{eff}}(\tau) \equiv \left( q_{-2}^2 / q_{-1} \right) Q_{\max}(\tau)$$

For the round Gaussian distribution, Eq. (26, 26a), $q_{-1} = 0.58$, $q_{-2} = 0.55$, $q_{-3} = 0.52$, yielding $q_{-2}^2 / q_{-1} = 0.52$.

Thus, Eq. (25) for eigenvalues $\nu$ and eigenfunctions $\bar{y}$ is valid for arbitrary beam transverse distribution, shape of the longitudinal potential well and arbitrary longitudinal distribution $f(H)$. Even if the longitudinal and transverse distributions are not factorized, Eq. (25) is still valid after proper modifications of the functions $u(\tau)$ and $Q(\tau)$.

For any real eigenvalue $\nu$, Eq. (25) has two independent solutions, even and odd one. In general, these solutions tend to non-zero constants at the tails, $\lim(\bar{y}(\tau))_{\tau \to \infty} = \bar{y}(\infty)$, while their derivatives $\bar{y}'(\tau)$ tend to zero,

$$\bar{y}'(\tau) \cong \frac{Q_{\text{eff}}(\tau)}{Q_{\text{eff}}(0)} \cdot (a - b\tau), \quad (28)$$

with constants $a$ and $b = \nu \bar{y}(\infty)$ being determined by the eigenvalue $\nu$. Without boundary conditions, the spectrum of Eq. (25) is continuous, while any boundary condition would select a sequence of discrete eigenvalues. Does any boundary condition have to be required? Note that the strong space charge and rigid beam approximations fail at the bunch tails. Namely, these assumptions are violated at that longitudinal offset, where the function $\bar{y}(\Psi)$ cannot be considered as slow function of the space charge phase $\Psi$ (see Eqs. 20-24). This happens at $\tau = \tau_*$, where

$$\left| \frac{d}{d\Psi} \bar{y}'(\tau_*) \right| \cong \left| \bar{y}'(\tau_*) \right|.$$

Using the asymptotical behavior of $\bar{y}'(\tau)$, Eq. (28), and the definition of the space charge phase $\Psi$ (Eq. 20), this yields an equation for that model-break point $\tau_*$ at the bunch tail:

$$Q(\tau_*) = u(\tau_*) | Q'(\tau_*) | / Q(\tau_*). \quad (29)$$

Here we assumed that at the tail $\bar{y}''/\bar{y}' \cong Q'/Q$. At this model-breaking point, the rigid-slice approximation fails. The individual particles, being essentially in coherent motion before that, go incoherently after that. As a result, the coherent motion longitudinally goes down much faster than it would go according to Eq. (28), were the model valid there. Note that it is not the coherent amplitude $\bar{y}(\tau)$ but its longitudinal derivative $\bar{y}'(\tau)$, what goes down at $\tau \geq \tau_*$. Indeed, the constant part of the amplitude $\bar{y}(\infty)$ does not decohere. A reason for that is the same as for the 0-th mode, $\nu=0$, $\bar{y}(\tau) = \text{const}$, which is an exact solution of the original equation of motion (1). Thus, we are coming to a conclusion, that at the model-breaking point $\tau_*$ the derivative of the eigenfunction goes to zero due to the fast decoherence. Thus, we are coming to the boundary condition:

$$\bar{y}'(\pm \tau_*) = 0. \quad (30)$$

This boundary condition is identical to what would be required if there were a vertical potential barrier at the model-breaking point. This additional meaning of the boundary condition (30) appears to be reasonable by itself. Indeed, setting that barrier at the model-breaking point makes the model applicable everywhere. At the same time, since it is set at that far tails, it almost does not change the collective dynamics of the bunch. The idea of model breaking implies that the right-hand side of Eq. (29) is defined up to a numerical factor ~ 1. However, since at the far tails the left-hand side of Eq. (29) is extremely fast function of its argument, the model-breaking point $\tau_*$ is defined with rather good accuracy at strong space charge. For instance, for Gaussian bunch with $Q_{\max}(0)/Q_s = 10$, reduction the right-hand side of Eq. (29) by a factor of 2 makes only 20% difference for $\tau_*$.

Eqs (25, 30) reduce the general problem of eigen-modes to a well-known mathematical boundary-value problem, similar to the single-dimensional Schrödinger equation (see e.g. [10]). This problem is normal, so it has full orthonormal basis of the eigen-functions at the interval $(-\tau_*, \tau_*)$.

$$\int_{-\tau_*}^{\tau_*} \bar{y}_k(\tau) \bar{y}_m(\tau) \frac{d\tau}{u(\tau)} = \delta_{km}.$$

As a consequence,

$$\sum_{m=0}^\infty \bar{y}_m(\tau) \bar{y}_m(s) = u(\tau) \delta(s - \tau).$$

At the bunch core, the *k*-th eigen-function $\bar{y}_k(t)$ behaves like $\sim \sin(k\tau/\sigma)$ or $\sim \cos(k\tau/\sigma)$, and the eigenvalues are estimated to be

$$\nu_k \cong k^2 \bar{Q}_s^2 / Q_{\text{eff}}(0) << \bar{Q}_s, \quad (31)$$

which are similar to the values in the square well case. With that orthogonality condition, the formulas for the coherent tune shift and the coherent detuning, Eqs (17,18) have to be generally modified by a substitution $d\tau \to d\tau/u(\tau)$:

$$Q_w = \kappa \iint_{-\infty\,\tau}^{\infty\,\infty} W(\tau-s) \exp(i\zeta(\tau-s))\rho(s)\bar{y}_k(s)\bar{y}_k(\tau)u^{-1}(\tau)dsd\tau,$$
$$Q_d = \kappa \iint_{-\infty\,\tau}^{\infty\,\infty} D(\tau-s)\rho(s)\bar{y}_k^2(\tau)u^{-1}(\tau)dsd\tau. \quad (32)$$

From here, it follows that a sum of all growth rates is zero:

$$\sum_{k=0}^{\infty} \text{Im} Q_w = 0. \quad (33)$$

This statement sometimes is referred to as the growth rates sum theorem. Note also that the detuning wake does not introduce any growth rate, and every growth rate is proportional to the head-tail phase when this phase is small, similar to the conventional no-space-charge case.

For a short wake, $W(\tau) = -G\delta'(\tau)$, the growth rate can be expressed as

$$\text{Im} Q_w = \kappa \rho_k \zeta G; \quad \rho_k \equiv \int_{-\infty}^{\infty} \rho(s)\bar{y}_k^2(\tau)u^{-1}(\tau)d\tau,$$

in agreement with the special result for a square well found in Ref. [6]. The same sign of the rates for all the modes may seem to contradict to the theorem (33). The contradiction is resolved when short wavelength modes are taken into account. Namely, the wake function cannot be approximated by the Dirac function for so short waves whose length is smaller than a scale of the wake function. These short waves introduce the required opposite sign contribution to the sum of the rates, making it zero.

Let's consider now an alternative case of a slowly decaying wake, that is we assume that $W(\tau) \approx -W_0 = \text{const}$. At small head-tail phases, the lowest mode has the same sign as the short-wake rate (34)

$$\text{Im} Q_{w0} \cong 0.4\kappa N_b \zeta \sigma W_0; \quad \zeta\sigma \leq 1$$

where $\sigma$ is the rms bunch length and $N_b$ is a number of particles in the bunch. The growth rates of the higher modes are of opposite sign, making the rate sum equal to zero, Eq. (33). As a function of chromaticity, the growth rates reach their maxima at $\zeta\sigma \cong 0.7(k-1), \quad k \geq 2$, where $\max(\text{Im} Q_w) \cong 0.1 \kappa N_b W_0$. After its maximum, the high order mode changes its sign at $\zeta\sigma \cong 0.7k, \quad k \geq 2$ to the same sign as the lowest mode, tending after that to

$$\text{Im} Q_w \cong \kappa N_b W_0 /(4\zeta\sigma). \quad (34)$$

All the numerical factors were estimated using the data of Fig. 3, showing the coherent rates for the Gaussian bunch and constant wake function (see the next section), which are also not too far from the square well results of Fig.1.

## 5. MODES FOR GAUSSIAN BUNCH

The Gaussian distribution in phase space,

$$f(v,\tau) = \frac{N_b}{2\pi \sigma u}\exp(-v^2/2u^2 - \tau^2/2\sigma^2), \quad (35)$$

deserves a detailed consideration as a good example of solving the general problem, and due to its special practical importance. Indeed, this distribution function describes a thermal equilibrium of a bunch whose length is much shorter than the RF wavelength. Below, natural units for this problem are used. The distance $\tau$ is measured in units of the bunch length $\sigma$, and the eigenvalue $\nu$ - in units of $u^2/(\sigma^2 Q_{\text{eff}}(0)) = Q_s^2/Q_{\text{eff}}(0)$.

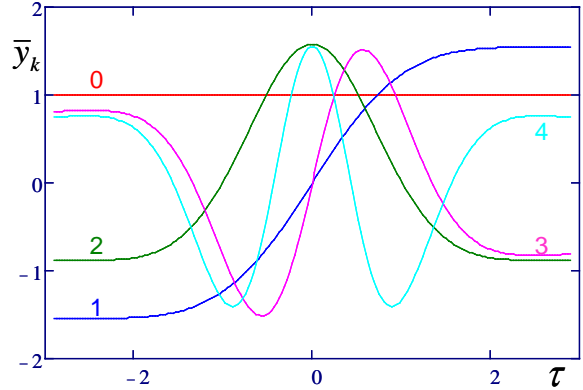

Fig. 2: The first five eigenfunctions for the Gaussian beam at $\tau_* = 2.5$ (or $q=60$) as functions of the dimensionless distance along the bunch $\tau$, Eq. (36). The eigenfunctions are identified by their mode numbers.

In these units, the boundary-value problem of Eqs. (25, 30) is written as

$$\nu_k \bar{y} + \frac{d}{d\tau}\left(e^{\tau^2/2}\frac{d\bar{y}}{d\tau}\right) = 0;$$
$$\bar{y}'(\pm \tau_*) = 0; \quad (36)$$
$$\tau_* = \sqrt{2\ln(q/\tau_*)}; \quad q \equiv Q_{\text{eff}}(0)/Q_s.$$

This equation is easily solved numerically. A list of first ten eigenvalues $\nu_k$ found for $\tau_*$=1.5, 2.0 and 2.5 (corresponding to $q$=5, 15 and 60) is presented in the Table 1.

Table 1: First ten eigenvalues $\nu_k$ of the Gaussian bunch (Eq. 36) for $\tau_*$=1.5, 2.0, 2.5.

| $\tau_* \backslash k$ | 0 | 1 | 2 | 3 | 4 | 5 | 6 | 7 | 8 | 9 |
|---|---|---|---|---|---|---|---|---|---|---|
| 1.5 | 0 | 1.2 | 5.2 | 11 | 19 | 28 | 40 | 53 | 68 | 84 |
| 2.0 | 0 | 0.78 | 4.0 | 9.2 | 17 | 26 | 37 | 50 | 65 | 81 |
| 2.5 | 0 | 0.55 | 3.2 | 7.7 | 14 | 22 | 32 | 45 | 60 | 75 |

All these eigenvalues are limited as

$$k^2 - k/2 \leq \nu_k \leq k^2 + k/2; \quad k = 0,1,2... \quad (37)$$

These numbers are only logarithmically sensitive to the space charge parameter $q >> 1$. The first four eigenfunctions, normalized to the unit 'energy',

$$\frac{1}{\sqrt{2\pi}}\int_{-\infty}^{\infty}\bar{y}_k^2(\tau)e^{-\tau^2/2}d\tau = 1, \qquad (38)$$

are shown in Fig. 2.

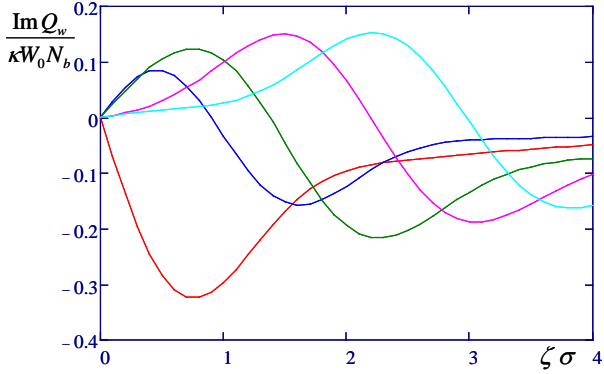

Fig. 3: Coherent growth rates for the Gaussian bunch with the constant wake as functions of the head-tail phase $\zeta\sigma$, for the lowest mode 0 (red), mode 1 (blue), 2 (green), 3 (magenta) and 4 (cyan). The rates are in units of $\kappa N_b W_0$, similar to the square well case of Fig. 1.

With the modes of the Gaussian bunch found, the coherent growth rates can be calculated according to Eq. (32), see Fig. 3. These growth rates of the Gaussian bunch look rather similar to the rates for the square well presented in Fig.1.

## 6. INTRINSIC LANDAU DAMPING

A goal for this section is to discuss a mechanism of Landau damping of the head-tail modes and calculate its rates at strong space charge. Both longitudinal RF and transverse lattice focusing are assumed linear here; that is why the discussed mechanism is called intrinsic, meaning that damping happens due to the space charge itself. Landau damping due to the lattice nonlinearity is calculated in the next section. The damping rates are found approximately, the accuracy is limited by a numerical factor ~ 1. The calculations are carried out for a Gaussian bunch, but the same method can be applied for any 3D distribution.

Landau damping is a mechanism of dissipation of coherent motion due to transfer of the energy into incoherent motion. The coherent energy is effectively transferred only to resonant particles - the particles whose individual frequencies are in resonance with the coherent motion. For these particles to exist, the incoherent spectrum must not be line, but continuous. How can these conditions be satisfied for a bunched beam with strong space charge? It may seem, at first glance, that when the space charge tune shift highly exceeds the synchrotron tune, the incoherent frequencies are so distant from the coherent line that the resonant particles do not exist at all, making the Landau damping impossible. This conclusion is not correct because it does not take into account the fact that the space charge tune shift is not constant along the bunch: being maximal at the bunch center, and dropping to zero at the tails. Thus, the Landau energy transfer is impossible at the bunch center, but it gets effective at the bunch tails, where the local incoherent space charge tune shift becomes small enough. Namely, for any given particle it happens at that distance, where a drop of the coherent amplitude is already so fast, that this drop is not adiabatic, and so energy of the coherent motion is transferred to the incoherent motion of this particle. Thus, for a given location τ, this energy transfer occurs for those particles whose velocities $v_i$ and individual space charge tune shifts $Q_i(\tau)$ relate as

$$Q_i(\tau) \cong v_i/\Delta\tau = |v_i Q'(\tau)|/Q(\tau) = v_i \tau. \qquad (39)$$

This equation is similar to Eq. (29), except that applied to the individual tail particle at the given location τ, not to the core of them, as Eq. (29). The individual local space charge tune shift in Eq. (39) is a function of the two transverse actions:

$$Q_i(\tau) \equiv Q(J_1, J_2, \tau). \qquad (40)$$

Using that the particle's longitudinal offset and velocity relate to its longitudinal action

$$J_\| = \frac{\tau^2}{2} + \frac{v^2}{2}, \qquad (41)$$

Eqs. (39-41) define at the given longitudinal position τ a 2D Landau surface in the space of three actions $J_1, J_2, J_\|$:

$$J_\| = \frac{\tau^2}{2} + \frac{Q^2(\tau, J_1, J_2)}{2\tau^2}, \qquad (42)$$

After passing its 'Landau point' (39), the particle gets the variable part of the coherent amplitude

$$\tilde{y}(\tau) \cong \bar{y}'(\tau)Q(\tau)/|Q'(\tau)| \qquad (43)$$

as its incoherent amplitude. For the Gaussian distribution

$$\tilde{y}(\tau) = \bar{y}'(\tau)/\tau. \qquad (43a)$$

Before proceeding with the estimate, the mentioned requirement for the continuous incoherent spectrum should be considered. Can this condition be satisfied without longitudinal or transverse non-linearity of the RF and the lattice? Contrary to the no-space-charge case, it can, because the betatron phase advance per the synchrotron period depends on the three actions. For the Gaussian bunch, the space charge phase advance for a particle with the longitudinal amplitude $\tau_0 = \sqrt{2J_\|}$ per the synchrotron period $T_s = 2\pi/Q_s$ is calculated as

$$\Psi_s(\tau_0) \equiv 4\int_0^{T_s/4} Q(\tau_0 \sin(Q_s\theta))d\theta =$$

$$\Psi_s(0)\exp\left(-\frac{\tau_0^2}{4}\right)I_0\left(\frac{\tau_0^2}{4}\right) \sim \sqrt{\frac{2}{\pi}}\frac{\Psi_s(0)}{\tau_0}.$$

This shows that the individual spectrum of particles at is indeed continuous. Since

$$\Psi_s(0)/(2\pi) = Q(0)/Q_s \gg 1,$$

there are many lines of the resonant particles, numbered by integer n, for which

$$\Psi_s(\tau_0) \equiv 2\pi n .$$

After M times of passing its Landau point, the particle gets its individual amplitude excited by

$$\Delta y_i(M) = \tilde{y} \sum_{m=0}^{M-1} e^{im\psi} = \tilde{y} e^{iM\psi/2} \frac{\sin(M\psi/2)}{\sin(\psi/2)}.$$

Thus, the entire Landau energy transfer for the bunch after $M \gg 1$ turns can be expressed as

$$\Delta E(M) = 4 \int d\mathbf{J} f(\mathbf{J}) \tilde{y}^2 \frac{\sin^2(M\psi/2)}{\sin^2(\psi/2)},$$

where $\mathbf{J}$ is 3D vector of the three actions, $\tilde{y} = \tilde{y}(\tau)$ is defined by Eq. (43, 43a), $J_\parallel = J_\parallel(\tau, J_1, J_2)$ is given by Eq.(42), so the 3D integral over actions has to be understood as

$$\int d\mathbf{J}(...) = \int_0^\infty dJ_1 \int_0^\infty dJ_2 \int_0^\infty d\tau \frac{\partial J_\parallel}{\partial \tau}(...);$$

leading and trailing bunch tails are taken into account. The contributions from particle entering and leaving the tails are assumed equal in magnitude but with random relative phase. The power of the Landau energy transfer is calculated as

$$\Delta \dot{E} = \frac{d\Delta E(M)}{T_s dM} = 4Q_s \int d\mathbf{J} f(\mathbf{J}) \tilde{y}^2 \delta_P(\psi),$$

$$\delta_P(\psi) \equiv \sum_n \delta(\psi - 2\pi n).$$

Here, it is used that at $M \gg 1$, $\sin(M\phi)/\phi = \pi \delta(\phi)$. Since the space charge phase advance $\Psi$ is a big number, the sum over many resonance lines $n$ can be approximated as an integral over these resonances, leading to

$$\Delta \dot{E} = \frac{2Q_s}{\pi} \int d\mathbf{J} f(\mathbf{J}) \tilde{y}^2. \quad (44)$$

This energy dissipation is directly related to the Landau damping rate $\Lambda_k$ by $\Delta \dot{E} = 2\Lambda_k E_k$, with the energy number $E_k$ given by

$$E_k = \left[ \int_{-\infty}^\infty Q(\tau) d\tau \right]^{-1} \int_{-\infty}^\infty \bar{y}_k^2(\tau) Q(\tau) d\tau.$$

This leads to a general formula for the Landau damping rate, valid for any kind of bunch 3D distribution:

$$\Lambda_k = \frac{Q_s}{\pi} \frac{\int_{-\infty}^\infty Q(\tau) d\tau}{\int_{-\infty}^\infty \bar{y}_k^2(\tau) Q(\tau) d\tau} \int d\mathbf{J} f(\mathbf{J}) \tilde{y}_k^2. \quad (45)$$

For the longitudinal Gaussian distribution, assuming the eigenfunctions normalized by the unit energy, as they were calculated in the previous section, it yields

$$\Lambda_k = \frac{Q_s}{\pi \tau_*^2} \int d\mathbf{J} f(\mathbf{J}) y_k'^2. \quad (45a)$$

Here the term $1/\tau^2$ was taken out from the integral and substituted by its decoherence (model-breaking) value $1/\tau_*^2$. According to Eqs. (28, 30),

$$y' \cong b \cdot (\tau - \tau_*) \frac{Q_{\text{eff}}(\tau)}{Q_{\text{eff}}(0)} = b \cdot (\tau - \tau_*) \exp(-\tau^2/2), \quad (46)$$

where the asymptotic parameter $b$ can be calculated for every eigenmode, $b=b_k$. For the modes of the Gaussian bunch, described in the previous section, numerically found squares of these parameters are presented in the Table 2. The Landau damping rate is proportional to that parameter, $\Lambda_k \propto b_k^2$; thus, it may be concluded that the damping rate is extremely sensitive to the mode number. Since the number of the lowest potentially unstable mode is about the chromatic head-tail phase (see Fig. 3), an increase of the chromaticity has to be a powerful tool for the beam stabilization.

Table 2: Mode asymptotic parameters $b_k^2$ for the modes of Table 1.

| $\tau_* \backslash k$ | 0 | 1 | 2 | 3 | 4 | 5 | 6 | 7 | 8 | 9 |
|---|---|---|---|---|---|---|---|---|---|---|
| 1.5 | 0 | 2.5 | 22 | 70 | 130 | 240 | 320 | 400 | 550 | 700 |
| 2.0 | 0 | 1.3 | 15 | 64 | 160 | 320 | 550 | 770 | 1000 | 1400 |
| 2.5 | 0 | 0.85 | 7.5 | 40 | 105 | 260 | 500 | 1060 | 1700 | 2300 |

The longitudinal integration in Eq. (45a) can be taken by the saddle-point method. After that, the remaining transverse integral leads to the form-factor $q_{-3}^{-3} \approx q_{-2}^2 / q_{-1} = 0.5$, see Eq.(27). In the result, the Landau damping rate is found as

$$\Lambda_k = 1.5 b_k^2 Q_s / q^3; \quad q = \frac{Q_{\text{eff}}(0)}{Q_s}. \quad (47)$$

Note that the synchrotron tune and the space charge tune shift enter in high powers in Eq. (47).

In this section, Landau damping was calculated assuming there is no nonlinearity in the lattice and the longitudinal force from the RF field. That is why that kind of Landau damping can be called intrinsic. Although the method of calculation is general, the specific results assume small coherent tune shift between the two neighbor modes, $Q_w \ll Q_s^2 / Q_{\text{eff}}(0)$. In particular, independence of the Landau damping rate of the chromaticity should not be expected for larger wake terms (32), since these terms directly depend on the chromaticity. For that large wake forces, the chromaticity changes the asymptotic of eigenfunctions, and thus, the Landau damping would change as well.

## 7. LANDAU DAMPING BY LATTICE NONLINEARITY

In the previous section, the Landau damping was estimated for a linear lattice, where the bare tunes do not depend on the transverse amplitudes. If it is not so, the lattice tune spectrum is continuous, which may contribute an additional part to the entire Landau damping. This contribution is considered in this section.

To begin, let $\delta Q(J_2)$ be a nonlinear correction to the individual betatron tune for the 1$^{st}$ degree of freedom, which depends only on another action $J_2$ and does not depend of its own action $J_1$. Possible dependence on $J_1$ is taken into account later on. The nonlinearity modifies the single-particle equation of motion Eq. (4) as

$$\dot{y}_i(\theta) = iQ(\tau_i(\theta))[y_i(\theta) - \bar{y}(\theta, \tau_i(\theta))] - \delta Q_i y_i(\theta) \quad (48)$$

For positive non-linearity, $\delta Q_i > 0$, there is a certain point in the bunch, $\tau = \tau_r$, where the nonlinear tune shift exactly

compensates the local space charge tune shift:
$$\delta Q_i = Q(\tau_r) \quad (49)$$
When the particle crosses this point, it crosses a resonance of its incoherent motion with the coherent one. Crossing the resonance excites the incoherent amplitude. Indeed, a solution of Eq. (48) is expressed similar to Eq. (20):

$$y_i(\theta) = -i\int_{-\infty}^{\theta} Q(\tau_i(\theta'))\bar{y}(\theta',\tau_i(\theta'))\exp(i\Psi(\theta) - i\Psi(\theta'))d\theta' ;$$
$$\Psi(\theta) \equiv \int_0^{\theta}[Q(\tau_i(\theta')) - \delta Q]d\theta' . \quad (50)$$

Taking the integral (50) by the saddle-point method yields the incoherent amplitude excitation by the resonance crossing:

$$|\Delta y_i| = Q(\tau_r)\bar{y}(\tau_r)\sqrt{\frac{2\pi}{|\dot{Q}(\tau_r)|}} . \quad (51)$$

Here
$$\dot{Q}(\tau_r) = \frac{dQ}{d\tau}\frac{d\tau}{d\theta} = \frac{dQ}{d\tau}v$$
is a time derivative of the local space charge tune shift in the resonance point (49) seen by the particle. For the Gaussian bunch, $\dot{Q} = -QQ_s\tau v$, in the dimensionless units of length $\tau$ and velocity

$$v = \sqrt{2(J_\| - J_r)} ; \quad J_r \equiv \tau_r^2/2 . \quad (52)$$

After the single-pass incoherent excitation is found, the multi-pass summation for the coherent energy dissipation can be done exactly as in the previous section, leading to an analogue of Eq. (44):

$$\Delta\dot{E} = \frac{2Q_s}{\pi}\int d\mathbf{J}_\perp \int_{J_r}^{\infty} dJ_\| f(\mathbf{J}_\perp, J_\|)|\Delta y_i|^2 , \quad (53)$$

with $\mathbf{J}_\perp = (J_1, J_2)$ as 2D transverse action. Like in the previous section, using $\Delta\dot{E} = 2\Lambda_k E_k$, this energy dissipation gives the Landau damping rate

$$\Lambda_k = \frac{Q_s}{\pi E_k}\int d\mathbf{J}_\perp \int_{J_r}^{\infty} dJ_\| f(\mathbf{J}_\perp, J_\|)|\Delta y_i|^2 , \quad (54)$$

Up to this point, it is assumed that the nonlinear tune shift is independent of the action $J_1$ associated with the considered plane of the oscillation. If this dependence exists, the result has to be modified similar to conventional head-tail case (no-space charge), $f \to -J_1 \partial f/\partial J_1$, leading to a general formula:

$$\Lambda_k = -\frac{Q_s}{\pi E_k}\int d\mathbf{J}_\perp \int_{J_r}^{\infty} dJ_\| \frac{\partial f(\mathbf{J}_\perp, J_\|)}{\partial J_1}J_1|\Delta y_i|^2 . \quad (55)$$

Eq. (55) is valid for any distribution function and arbitrary nonlinearity.

For a Gaussian bunch (35), with the space charge tune shift (26), and symmetrical nonlinearity $\delta Q = \langle\delta Q\rangle \cdot (J_1 + J_2)/2$, evaluating the integral leads to

$$\Lambda_k = A\frac{\bar{y}_k^2(\infty)}{\tau_*}\frac{\langle\delta Q\rangle^2}{Q_{\max}(0)} ; \quad A \approx 1\cdot 10^2 \quad (56)$$

$$\bar{y}_k(\infty) = \begin{cases} 1, & \text{if } k = 0 \\ -b_k/\nu_k, & \text{otherwise} \end{cases}$$

where and $\langle\delta Q\rangle$ is an average value for the nonlinear tune shift.

Note that sign of the nonlinear tune shift is crucial for the Landau damping. In this section, positive sign of nonlinearity is assumed. For that case, particles with higher transverse amplitudes have additional positive tune shift. For high enough amplitudes, this compensates their negative tune shift from the space charge, opening a door for the Landau damping, calculated in this section. For negative nonlinear tune shift, this compensation does not happen; thus, negative nonlinearity should be either useless or detrimental for the beam stabilization.

A complete Landau damping rate is a sum of the two rates: the intrinsic rate found in the previous section and the nonlinearity-related rate of Eq. (55).

## 8. VANISHING TMCI

When the wake-driven coherent tune shift is small compared with the distance between the modes, $Q_w \ll Q_s^2/Q_{\max}$, it is sufficient to take it into account in the lowest order, as it is done by Eq. (32). When the wake force is not so small, that approximation is not justified. In this case, the modes are strongly perturbed by the wake forces; their shapes and frequencies significantly differ from the no-wake solutions of Eq. (25).

To complete the theory of the space charge modes, this equation has to be generalized for arbitrary wake field. To do so the wake term has to be dealt with in the same way as the space charge term, $\propto \bar{y}$ was dealt.

If the coherent tune shift is small compared with the space charge tune shift, the generalized equation for the eigen-modes follows:

$$\nu\,\bar{y}(\tau) + u(\tau)\frac{d}{d\tau}\left(\frac{u(\tau)}{Q_{\text{eff}}(\tau)}\frac{d\bar{y}}{d\tau}\right) = \kappa\left(\hat{\mathbf{W}}\bar{y} + \hat{\mathbf{D}}\bar{y}\right)$$

$$\hat{\mathbf{W}}\bar{y} \equiv \int_\tau^\infty W(\tau-s)\exp(i\zeta(\tau-s))\rho(s)\bar{y}(s)ds ; \quad (57)$$

$$\hat{\mathbf{D}}\bar{y} \equiv \bar{y}(\tau)\int_\tau^\infty D(\tau-s)\rho(s)ds .$$

Note that this wake-modified equation is valid for any ratio between the coherent tune shift and the synchrotron tune; only small value of the coherent tune shift compared with the space charge tune shift, $Q_w \ll Q_{\max}$, is required.

A straightforward way to solve Eq. (57) is to expand the eigenfunction $\bar{y}(\tau)$ over the full orthonormal basis of the no-wake modes $\bar{y}_{0k}(\tau)$:

$$\bar{y}(\tau) = \sum_{k=0}^{\infty} B_k \bar{y}_{0k}(\tau), \quad (58)$$

with as yet unknown amplitudes $B_k$. After that, the integro-differential Eq. (57) is transformed into a linear matrix problem for eigen-solutions

$$(\kappa\hat{\mathbf{W}} + \kappa\hat{\mathbf{D}} + \text{Diag}(\mathbf{v}_0))\mathbf{B} = \nu\mathbf{B} . \quad (59)$$

Here $\hat{\mathbf{W}}$ and $\hat{\mathbf{D}}$ are the matrices of the driving and detuning wake operators in the basis of the no-wake

modes:

$$\hat{\mathbf{W}}_{km} = \int_{-\infty}^{\infty}\int_{\tau}^{\infty} W(\tau-s)\exp(i\zeta(\tau-s))\rho(s)\bar{y}_{0k}(\tau)\bar{y}_{0m}(s)u^{-1}(\tau)dsd\tau , \quad (60)$$

$$\hat{\mathbf{D}}_{km} = \int_{-\infty}^{\infty}\int_{\tau}^{\infty} D(\tau-s)\rho(s)\bar{y}_{0k}(\tau)\bar{y}_{0m}(\tau)u^{-1}(\tau)dsd\tau ;$$

a symbol $\mathrm{Diag}(\mathbf{v}_0)$ represents a diagonal matrix whose $k$-th diagonal element is a $k$-th eigenvalue of the no-wake problem $v_{0k}$:

$$\mathrm{Diag}(\mathbf{v}_0)_{mn} = v_{0m}\delta_{mn};$$

and $\mathbf{B}$ is a vector of the mode amplitudes in Eq. (58).

In the conventional no-space-charge head-tail theory (see e. g. Ref.[4]), the beam is stable at zero chromaticity and the wake amplitude below a certain threshold. When the wake term grows, it normally moves the tunes $v_k = kQ_s$, $k = 0,\pm 1,\pm 2,...$ increasingly down. Mostly the tune of the mode number 0, or the base tune, is moved. As a result, that base tune meets at some threshold intensity the nearest from below tune of the mode -1, which is typically moved not as much. Starting from this point, the transverse mode coupling instability (TMCI) occurs. The threshold value of the coherent tune shift is normally about the synchrotron tune.

There is a significant structural difference between the conventional synchro-betatron modes and the space charge modes introduced here. The conventional modes are numbered by integer numbers, both positive and negative, while the space charge modes are numbered by positive integers only. All the tunes of the space charge modes are positive (i.e. they are above the bare betatron tune), increasing quadratically with the mode number. Due to fundamental properties of wake functions, the tunes are normally moved down by the wake force, and normally the mode 0 is affected most. At this point an important difference between the conventional and the space charge modes appears. Namely, in the conventional case, the 0 mode has a neighbor with lower tune. On the contrary, the space charge lowest mode does not have a neighbor from below; thus, its downward shifted tune cannot cross a tune of some other mode. Moreover, since the wake-driven tune shift normally decreases with the mode number, the wake field works as a factor of divergence of the coherent tunes. As soon as this typical picture is valid, TMCI is impossible. An illustration of this mode behavior is presented in Fig.4, where coherent tunes of the Gaussian bunch are shown as functions of the wake amplitude for a case of a constant wake $W = -W_0 = \mathrm{const} < 0$, no detuning, and zero chromaticity. The eigenvalues are given in the same units as they were calculated in the no-wake case and presented in Table 1.

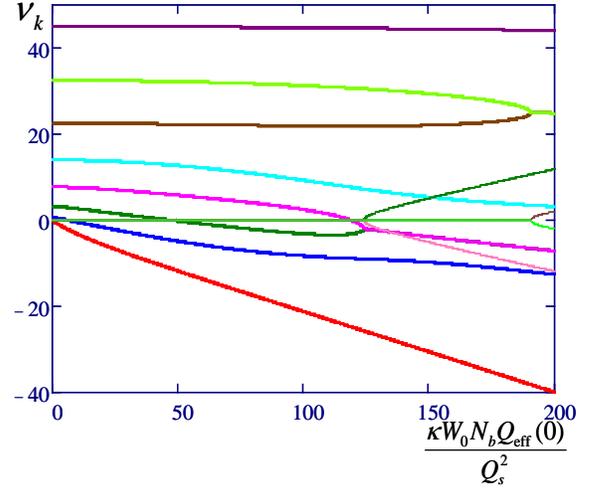

Fig.4: Coherent tunes of the Gaussian bunch for zero chromaticity, constant wake and no detuning, versus the wake amplitude. The eigenvalues are in the units of Table 1. Thick and thin lines show real and imaginary parts of the eigenvalues correspondingly. Colors of the real and imaginary parts are matched. Note high value of the TMCI threshold.

The TMCI still appears, as it is seen in Fig.4, but at very high values of the wake field, where formally calculated lowest order coherent tune shift is an order of magnitude higher than a tune separation between the lowest 2 modes. A similar mode behavior is shown in Fig. 5, for the resistive wall wake function, $W(\tau) = -W_0/\sqrt{\tau}$.

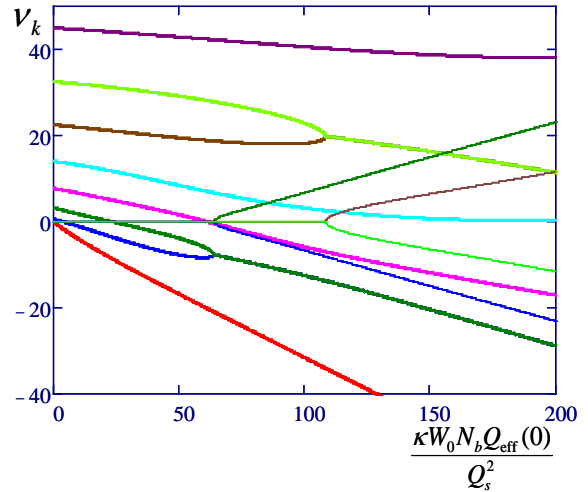

Fig.5: Same as Fig.4, but for the resistive wake function $W(\tau) = -W_0/\sqrt{\tau}$.

The conclusion about vanishing TMCI is also confirmed by calculations of Ref. [6] for an air-bag bunch in the square well with an exponential wake $W(-\tau) = -W_0 \exp(-\alpha\tau)$. In particular, Fig.14 of that article shows disappearance of TMCI for any wake length $1/\alpha$, as soon as the wake-driven coherent tune shift is exceeded by the space charge tune shift. The conclusion of this TMCI suppression at strong space charge should not be taken for granted for any wake function. TMCI suppression not necessarily happens for significantly oscillating wake functions, for which diagonal matrix elements have a maximum for one of the high-order modes. As a result, mode coupling would appear at rather low coherent tune shift for the oscillating wake functions. A mode behavior like that is illustrated in Fig. 6, where the coherent frequencies are calculated for the square potential well with an oscillating wake function $W(\tau) = -W_0 \cos(2\pi\tau/l)$. For this oscillating wake, the lowest $0^{th}$ mode is not shifted at first approximation, since its diagonal matrix element is zero due to the wake oscillations. The $1^{st}$ mode is the most affected by the wake field, being strongly shifted to the almost unaffected $0^{th}$ mode. As a result, the TMCI threshold is close to a point where it can be expected from the zero-wake slope of the $1^{st}$ mode; there is no any suppression of TMCI.

From practical point of view, wakes of hadron machines are typically dominated by the resistive wall contributions, which is constant-like in this context. In this case the TMCI threshold is significantly increased when the space charge tune shift exceeds the synchrotron tune.

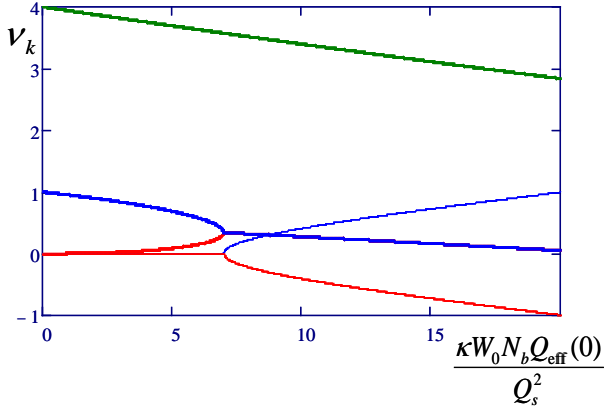

Fig. 6: Mode coupling for the square potential well and an oscillating wake function $W(\tau) = -W_0 \cos(2\pi\tau/l)$. Note that the TMCI threshold is where it can be expected from low-wake behavior of the most affected mode.

To summarize, an entire picture of the TMCI threshold can be described for arbitrary ratio of the space charge tune shift and the synchrotron tune. When this ratio is small, the conventional TMCI theory is applicable, giving approximately the synchrotron tune as the threshold value for the maximal coherent tune shift. When the space charge tune shift starts to exceed the synchrotron tune, the TMCI threshold for the coherent tune shift $(Q_w)_{th}$ is approximately determined by a minimum of two values: the space charge tune shift $Q_{max}$ and the lowest tune of the space charge mode $\sim Q_s^2/Q_{max}$, multiplied for non-oscillating wakes by a rather big numerical factor, 20-100. A very schematic picture of the TMCI threshold as a function of the space charge tune shift over the synchrotron tune is presented in Fig.7.

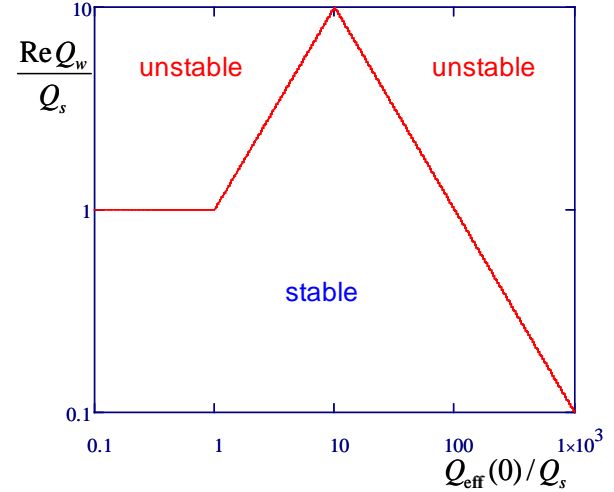

Fig.7: A schematic behavior of the TMCI threshold for the coherent tune shift versus the space charge tune shift. Both tune shifts are in units of the synchrotron tune.

This schematic picture would be additionally modified by multi-turn wake field or multiple bunches. This sort of coherent interaction can be taken into account for the space charge modes in a same manner it is done for the conventional head-tail modes [4]. Both for the conventional and the space charge modes, it is achieved by additional summation over bunches and the previous revolutions in the wake term.

## 9. SUMMARY

In this paper, a theory of head-tail modes is presented for space charge tune shift significantly exceeding the synchrotron tune, which is rather typical case for hadron machines. A general equation for the modes is derived for any ratio of the synchrotron tune and the wake-related coherent tune shift. Without the wake term, this is a 2-nd order self-adjoint ordinary differential equation, known to have full orthonormal basis of the eigenfunctions. The spectrum of this equation is discussed in general and solutions for the Gaussian bunch are presented in detail. Landau damping of the space charge modes is considered and calculated both without and with lattice nonlinearity. Finally, the transverse mode coupling instability for the space charge modes is considered. It is shown, that typically the TMCI threshold is 1-2 orders of magnitude higher than that naively expected from the small wake behavior of the lowest mode.

The presented theory needs to be compared with simulations and measurements. The author hopes this will happen in a near future.

## ACKNOWLEDGEMENTS

I am indebted to Valeri Lebedev, Oliver Boine-Frankenheim and Vladimir Kornilov for extremely useful discussions; my special thanks are to Norman Gelfand for his numerous linguistic corrections.

## REFERENCES

[1] C. Pellegrini, Nuovo Cimento 64A, p. 447 (1969)
[2] M. Sands, SLAC Rep. TN-69-8, TN-69-10 (1969).
[3] F. Sacherer, CERN Rep. SI-BR/72-5 (1972); F. Sacherer, Proc. 9$^{th}$ Int. Conf. High Energy Accel., SLAC, 1974, p. 374.
[4] A. Chao, "Physics of Collective Beam Instabilities in High Energy Accelerators", J. Wiley & Sons, Inc., 1993.
[5] "Handbook of Accelerator Physics and Engineering", ed. by A. W. Chao and M. Tigner, World Scientific,1998, p. 136, Eq. (22). Our Eq.(26) for the round beam follows after a substitution $a=1$, $1/(1+u)=z$. It has to be taken into account though that Eq. (22) of the "Handbook" contains a typo, confirmed by A. W. Chao in private communication: parameters $\alpha_{x,y}$ in that equation must be substituted by $\alpha_{x,y}/4$.
[6] M. Blaskiewicz, Phys. Rev. ST Accel. Beams, **1**, 044201 (1998).
[7] M. Blaskiewicz, Phys. Rev. ST Accel. Beams **6**, 014203 (2003).
[8] A. Burov, V. Lebedev, Phys. Rev. ST Accel. Beams 12, 034201 (2009).
[9] A. Burov, V. Danilov, Phys. Rev. Lett. **82**, 2286 (1999).
[10] E. Kamke, "Handbook of ordinary differential equations" (in German, 1959, or Russian translation, 1976).